# Chemical Relaxation of a Binary Mechanical Model System

*Josh E. Baker*

Department of Pharmacology, University of Nevada, Reno School of Medicine, Reno, Nevada

ABSTRACT

With potential relevance to biomechanics, an interesting problem in statistical mechanics not previously solved is a binary mechanical model system. Discrete chemical states of proteins are often associated with discrete metastable structural states, such that with a change in state a protein acts as a molecular switch. An ensemble of molecular switches that displace compliant elements equilibrated with an external force, $F$, constitutes a binary mechanical model system. As one in a series of publications developing this model, here I consider the mechanical performance of this system. Four processes naturally emerge from a transient analysis which are consistent with the four phases observed in a muscle force transient.

INTRODUCTION

Binary model systems are often presented as a basic introduction to the principles of statistical mechanics with the typical example being that of a system of $N$ spin ½ particles in one of two states, +½ and –½, where the distribution of states, $N_{+½}/N_{-½}$, varies exponentially with the strength of an applied magnetic field (1). An interesting parallel problem with potential relevance to biomechanics is a binary mechanical system of $N$ proteins in one of two states, AM and M, (Fig. 1A) where the distribution of states $N_{AM}/N_M$ varies exponentially with force, $F$. We previously developed the constant $F$ case (2). However, to describe a chemical relaxation within this system requires a more explicit definition of force generation. In a series of papers, we describe the mechanics, energetics, entropy, and kinetics of this system and show that it reconciles disparate models of molecular motor mechanochemistry (3). Here I consider the mechanical performance of this system.

Springs are useful constructs for describing reversible changes in force and energy associated with distortions of structural elements, molecules, or molecular assemblies (4, 5). However, a spring accurately describes the mechanical state of a molecule or assembly only if the energy within that spring is constrained, stored, and contained (like in a spring). Molecules within a thermodynamic system held at a constant temperature and pressure freely exchange energy with their surroundings, and thus molecular springs are defined thermally not mechanically. A thermodynamic system has only one mechanical constraint, and therefore only one spring that can be defined mechanically – a thermodynamic system spring. A thermodynamic spring in a binary mechanical model must describe reversible changes in system force and energetics in response to displacements generated both internally through state transitions and externally through changes in $F$. It must also be thermodynamically consistent, describing an exponential dependence of $N_{AM}/N_M$ on an applied force, $F$. The system spring illustrated in Fig. 1B (bottom) uniquely satisfies these requirements.

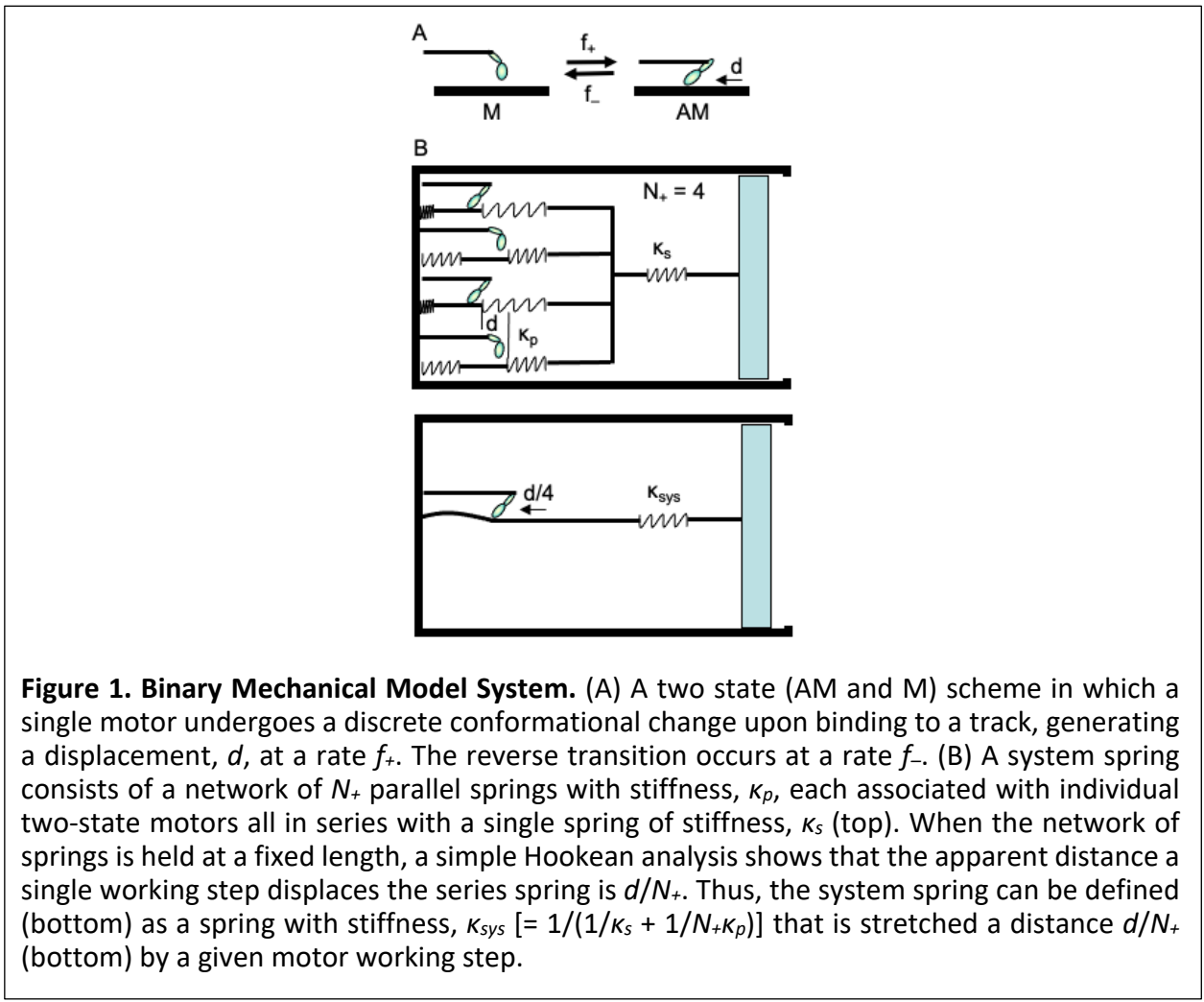

**Figure 1. Binary Mechanical Model System.** (A) A two state (AM and M) scheme in which a single motor undergoes a discrete conformational change upon binding to a track, generating a displacement, *d*, at a rate $f_+$. The reverse transition occurs at a rate $f_-$. (B) A system spring consists of a network of $N_+$ parallel springs with stiffness, $\kappa_p$, each associated with individual two-state motors all in series with a single spring of stiffness, $\kappa_s$ (top). When the network of springs is held at a fixed length, a simple Hookean analysis shows that the apparent distance a single working step displaces the series spring is $d/N_+$. Thus, the system spring can be defined (bottom) as a spring with stiffness, $\kappa_{sys}$ [= $1/(1/\kappa_s + 1/N_+\kappa_p)$] that is stretched a distance $d/N_+$ (bottom) by a given motor working step.

Using both a chemical thermodynamic analysis and computer simulations, here I show that four processes naturally emerge from a transient analysis of a binary mechanical system. This model applies to any system of molecular switches that displace compliant elements equilibrated with an external force. Here, I compare the model to a muscle system within which the molecular switches are myosin motors. Single molecule structural and mechanics studies of myosin provide direct evidence for the two states in Fig. 1A. The M to AM transition (a working step) generates force when a distinct conformational change in myosin (M) induced by strong binding to actin

(A) displaces compliant elements (6–12) a distance, $d$. We have demonstrated experimentally that the mechanical state of a myosin motor within an ensemble is not mechanically defined (13–17) and that the myosin working step equilibrates with muscle force (18), consistent with the thermodynamic model system in Fig. 1B. Here I show that the four phases that emerge from an analysis of a binary mechanical system accurately account for the four phases of a force transient observed in muscle.

Molecular mechanics defines molecular springs mechanically not thermally and provides the foundation for most models of molecular motor ensembles to date (5, 19). The application of thermodynamics to a motor ensemble is novel and provides a new perspective on how muscle and molecular motors work.

MATERIALS AND METHODS

A binary thermodynamic system consists of $N$ motors (molecular switches) that exist in one of two states (M and AM in Fig. 1A), where a transition from M to AM (a working step) displaces a distance $d/N_+$ a thermodynamic spring of stiffness $\kappa_{sys}$ performing work, $\langle F \cdot d \rangle = F \cdot d / N_+$, where $N_+ = y \cdot N$ is the number of force-generating motors (Fig. 1B), which is a fraction, $y$, of the total number, $N$, of motors. The free energy for the working step is

$$\Delta_r G = \Delta G^\circ + kT \cdot \ln(N_{AM}/N_M) + \langle F \cdot d \rangle \qquad \text{Eq. 1}$$

where $\Delta_r G$ is the Gibbs reaction free energy and $\Delta G^\circ$ is the Gibbs standard free energy.

The net forward flux through the working step,

$$dN_{AMD}/dt = -N_{AM} \cdot f_- + N_M \cdot f_+, \qquad \text{Eq. 2}$$

collective generates force, $\varkappa_{sys} \cdot d/N_+$, in the system spring at a rate

$$dF/dt = (dN_{AM}/dt) \cdot \varkappa_{sys} \cdot d/N_+,  \qquad \text{Eq. 3}$$

where $f_+$ and $f_-$ are the forward and reverse working step rate constants, and $N_{AM}$ and $N_M$ are the numbers of motors in the AM and M states.

Described in greater detail elsewhere (manuscript in review), a perturbation of the system through a rapid change in $F$ or $\Delta G^o$ changes the corresponding energy terms in Eq. 1 by $\delta E$. The perturbation also changes the equilibrium value for $kT \cdot \ln(N_{AM}/N_M)_{eq}$ by $\Delta G = \Delta_r G$. The ratio of the non-equilibrium forward, $f_+$, and reverse, $f_-$, rate constants can be derived from Eq. 1 to obtain

$$f_+/f_- = \exp[(-<F \cdot d> - \delta E - \Delta G - \Delta G^o)/kT]. \qquad \text{Eq. 4}$$

For each energy term, $E$, in Eq. 4, the $E$-dependence of $\exp(E/kT)$ can be partitioned between forward, $f_+(E)$, and reverse, $f_-(E)$, rate constants through a coefficient, $a_E$, that describes the fractional change in $E$ prior to the activation energy barrier as

$$f_+(E) = \exp(a_E E/kT) \text{ and}$$

$$f_-(E) = \exp[-(1 - a_E)E/kT).$$

If the chemical relaxation described by Eqs. 2 and 3 rapidly equilibrates with the surroundings, $\Delta G = \Delta_r G$ follows the time course of $N_{AM}$ and $N_M$ (Eqs. 2 and 3). If the chemical relaxation does not rapidly equilibrate with the surroundings [i.e., $\Delta G(t)$ is exchanged with the surroundings more slowly than the relaxation rate defined by Eqs. 2 and 3] then the chemical relaxation is adiabatic, and the exchange of $\Delta G(t)$ with the surroundings is damped as

$$d\Delta G/dt = -\Delta G(t) \cdot b \qquad \text{Eq. 5}$$

where *b* is the rate of equilibration.

**Computer Simulations.** Using master equations 2, 3, and 5, the rate constants defined by Eq. 4, and the model parameters in Table I, MatLab (Mathworks, Natick, MA) is used to simulate muscle force transients.

RESULTS

Reaction isotherms (Fig. 2, parabolas) describe changes in the free energy, G, of a system (y-axis) associated with changes in the extent, $\xi(N_{AM}, N_M, F)$, of the reaction (x-axis), where the slope of the isotherm is the reaction free energy, $\Delta_r G = (\partial G/\partial \xi)_{T,P}$. The connection between the free energy diagram in Fig. 2 and the kinetics and energetics described above is that Eqs. 2 and 3 describe the extent of the coupled reaction, $\xi(N_{AM}, N_M, F)$; Eq. 4 describes the kinetics and energetics of the coupled reaction, $\Delta_r G = (\partial G/\partial \xi)_{T,P}$; and Eq. 5 describes $\Delta G(t)$ when the coupled reaction is adiabatic and does not follow the reaction isotherm. The mechanical performance of a binary system is diagramed in Fig. 2 and simulated in Fig. 3.

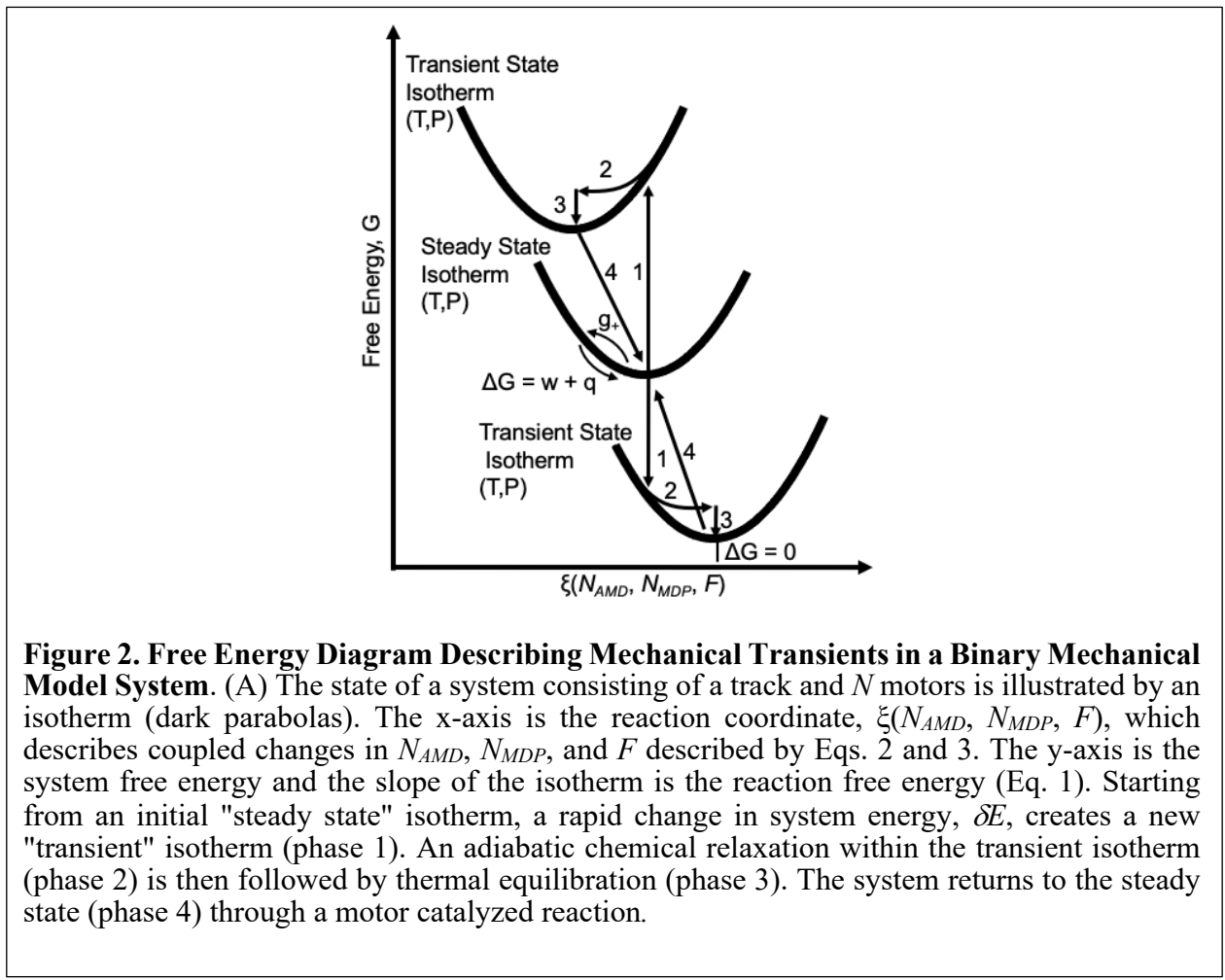

**Figure 2. Free Energy Diagram Describing Mechanical Transients in a Binary Mechanical Model System.** (A) The state of a system consisting of a track and $N$ motors is illustrated by an isotherm (dark parabolas). The x-axis is the reaction coordinate, $\xi(N_{AMD}, N_{MDP}, F)$, which describes coupled changes in $N_{AMD}$, $N_{MDP}$, and $F$ described by Eqs. 2 and 3. The y-axis is the system free energy and the slope of the isotherm is the reaction free energy (Eq. 1). Starting from an initial "steady state" isotherm, a rapid change in system energy, $\delta E$, creates a new "transient" isotherm (phase 1). An adiabatic chemical relaxation within the transient isotherm (phase 2) is then followed by thermal equilibration (phase 3). The system returns to the steady state (phase 4) through a motor catalyzed reaction.

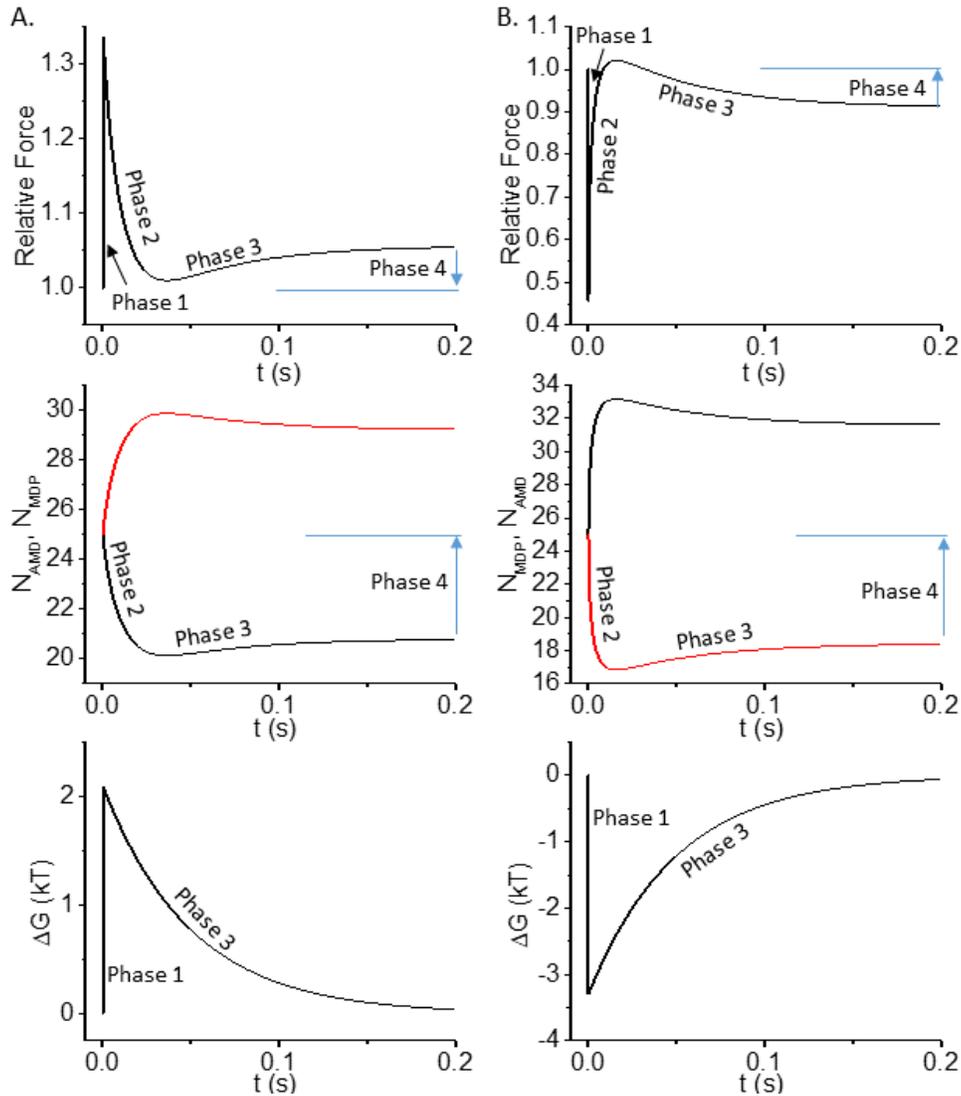

**Figure 3.** Computer simulations of a chemical relaxation following both a rapid increase (A) and decrease (B) in the length of the system spring. (A) At t = 0, $N_{AMD}$ and $N_{MDP}$ are set to 25 and relative force is normalized to $F_i$ = 138 pN to satisfy Eq. 1 ($\Delta G$ = 0). A system spring displacement of $\Delta L$ changes the system force from $F_i$ to $F_i + \Delta L \cdot \kappa_{eff}$ with a corresponding change in energy $\delta E = (<(F_i + \Delta L \cdot \kappa_{eff}) \cdot d> - <F_i \cdot d>)$. Phase 1 is simulated by setting $F$ to $F_i + \Delta L \cdot \kappa_{eff}$ and initializing $\Delta G(t)$ to $\Delta_r G$ in Eq. 4. A chemical relaxation is then simulated using master equations 2, 3, and 5 with kinetics and energetics defined by Eq. 4. Shown are the simulated time courses of relative force (top panel), $N_{AMD}$ (black) and $N_{MDP}$ (red) (center panel), and $\Delta G(t)$ (bottom panel). (A) Computer simulations following a positive $\Delta L$, with phases labeling the four dominant processes of perturbation (phase 1) chemical relaxation (phase 2), equilibration (phase 3), and return to initial "steady state" values (phase 4). (B) Computer simulations following a negative $\Delta L$, with the same four processes labeled.

**Phase 1**. Starting from equilibrium within a "steady state" isotherm, phase 1 occurs with a rapid change in system energy, $\delta E$, associated with a rapid change in $F$ or $\Delta G°$ (r.h.s. Eq. 1), creating a new "transient" isotherm (Fig. 2). A positive $\delta E$ (Fig. 2, up-arrow) results in a positive $\Delta_r G$ (Fig. 2, positive slope) which drives the force-generating working step reaction in the reverse direction. A negative $\delta E$ (Fig. 2, down-arrow) results in a negative $\Delta_r G$ (Fig. 2, negative slope) which drives the force-generating working step reaction in the forward direction. A spectrum of transient isotherms is created when different amounts of energy are added to or removed from the system, with each isotherm describing a unique relationship between $\Delta G$ and $\xi(N_{AM}, N_M, F)$ with unique equilibrium values for $N_{AM}$, $N_M$, and $F$.

Phase 1 is simulated by initializing model parameters in Eqs. 2 – 4 to equilibrium values and then setting the parameter being perturbed to its new value and $\Delta G(t)$ to $\Delta_r G$ (Eq. 1). Figure 3 shows simulations of phase 1 for both a rapid increase (positive $\Delta L$, Fig. 3A) and rapid decrease (negative $\Delta L$, Fig. 3B) in the system spring length, changing the force from $F_i$ to $F_i \pm \Delta L \cdot \kappa_{eff}$ with a corresponding change in energy $\delta E = <(F_i \pm \Delta L \cdot \kappa_{eff}) \cdot d> - <F_i \cdot d>$.

**Phase 2**. Phase 2 is the chemical response to the phase 1 perturbation and occurs down a transient isotherm (Fig. 2) in a direction that minimizes $G$. As illustrated in Fig. 2, a perturbation, $\delta E$, that either increases (phase 1, up arrow) or decreases (phase 1, down arrow) the energy of the system results in a chemical relaxation in a direction that reverses the effects of $\delta E$, satisfying Le Chatellier's principle. A chemical relaxation is non-adiabatic if it rapidly equilibrates with the surroundings. In this case, the chemical relaxation occurs with a single phase (phase 2) along the isotherm with $N_{AMD}$, $N_{MDP}$, and $F$ rapidly approaching values defined at $\Delta G = 0$. However, if the exchange of free energy, $\Delta G(t)$, with the surroundings is damped on the time scale of the chemical relaxation, the chemical relaxation is adiabatic, and the chemical relaxation overshoots the isotherm (phase 2 in Fig. 2) with $N_{AM}$, $N_M$, and $F$ approaching values defined at $\Delta G(t)$ not $\Delta G = 0$.

Figure 3 shows simulated values for $F$ (Fig. 3, top panel) and $N_{AM}$, and $N_M$ (Fig. 3, middle panel) following the phase 1 perturbation described above. The simulations exhibit two distinct phases (phase 2 and phase 3). Phase 2 is primarily determined by Eqs. 2 and 3 at a roughly constant $\Delta G(t)$. Because Eq. 4 is derived from the system free energy, the chemical relaxation inherently occurs in a direction that minimizes the system free energy and is consistent with Le Chatellier's principle, approaching values for $N_{AM}$, $N_M$, and $F$ defined at a non-equilibrated $\Delta G(t)$ that overshoot values defined at $\Delta G = 0$.

**Phase 3**. Following an adiabatic chemical relaxation (phase 2) the system eventually equilibrates (phase 3, Fig. 2) when $\Delta G(t)$ is exchanged with the surroundings and the system returns to the energy minimum of the transient isotherm ($\Delta G = 0$). This relatively slow change in $\Delta G(t)$ results in a relatively slow shift in values for $N_{AMD}$, $N_{MDP}$, and $F$ from those defined at $\Delta G(t)$ to those defined at $\Delta G = 0$. Thus, the non-equilibrated $\Delta G(t)$ that drove the chemical relaxation past equilibrated values in phase 2 is, in phase 3, exchanged with the surroundings upon equilibration, reversing non-equilibrated phase 2 gains.

In Fig. 3, phase 3 is primarily determined by Eq. 5. Changes in $N_{AM}$, $N_M$, and $F$ occur with changes in $\Delta G(t)$ (Eq. 5; Fig. 3, bottom panel) in a direction opposite phase 2 because as $\Delta G(t)$ approaches zero, the thermodynamically consistent kinetics in Eq. 4 change such that $N_{AMD}$, $N_{MDP}$, and F through Eqs. 2 and 3 approach equilibrated ($\Delta G = 0$) values. Note that $\Delta G = 0$ both before phase 1 and after phase 3 even though the corresponding equilibrium values for $N_{AMD}$, $N_{MDP}$, and $F$ are different. This demonstrates kinetically how a new isotherm is created upon adding or removing energy, $\delta E$, as described in Fig. 2 (phase 1).

The effects of both $b$ and the extent (0 – 100%) of the adiabatic relaxation are simulated in Fig. 4A, showing that $b$ influences the rate of phase 3 (Fig. 4A, inset) and that the extent of adiabatic relaxation influences the amplitude of phase 3. At either 0% adiabatic relaxation (bottom curve) or sufficiently fast $b$ (bottom curve, inset) phase 3 is eliminated.

**Phase 4**. In an active motor system, a motor-catalyzed reaction (20–22) irreversibly transfers motors from AM to M, adding energy $\delta E$ to the system at a relatively slow rate ($g_+$ in Fig. 1) which is then lost from the system as heat and work performed on the surroundings upon a chemical relaxation (down the isotherm). Over time ($1/g_+$), this repeated perturbation-equilibration cycle results in steady state values for $N_{AMD}$, $N_{MDP}$, and $F$ that are unique to a "steady state" isotherm. Thus, in an active motor system a fourth phase occurs with a transition from a transient isotherm back to the initial "steady state" isotherm at a rate $g_+$ and an amplitude equal to the difference between the transient and steady state equilibrated values.

Here, I compare the phases that are intrinsic to both an energetic (Eq. 2) and kinetic (Eq. 3) analysis of a binary mechanical model system with the phases observed in muscle force transient experiments.

**Length-Step Simulations**. A rapid change in the length of steady state isometric muscle results in a multi-phasic force response. This was first observed in whole muscle (23, 24) and subsequently observed with improved time resolution in single muscle fibers (25, 26). The latter studies reveal four distinct phases.

Figure 4B shows an overlay of simulated transient responses to length steps of different amplitudes. The transient responses are non-exponential, and phase 2 rates were determined from the simulations as $1/t_{½}$ where $t_{½}$ is the time at which the force reaches ½ the sum of the maximum phase 1 and phase 3 force. Figure 4C is a plot of simulated phase 2 rates for different length steps overlaid with experimental data obtained from R temporaria muscle at 2° C (25). Figure 4D is a plot of the simulated phase 2 amplitudes for different length steps overlaid with experimental data obtained from R temporaria muscle at 2° C (25).

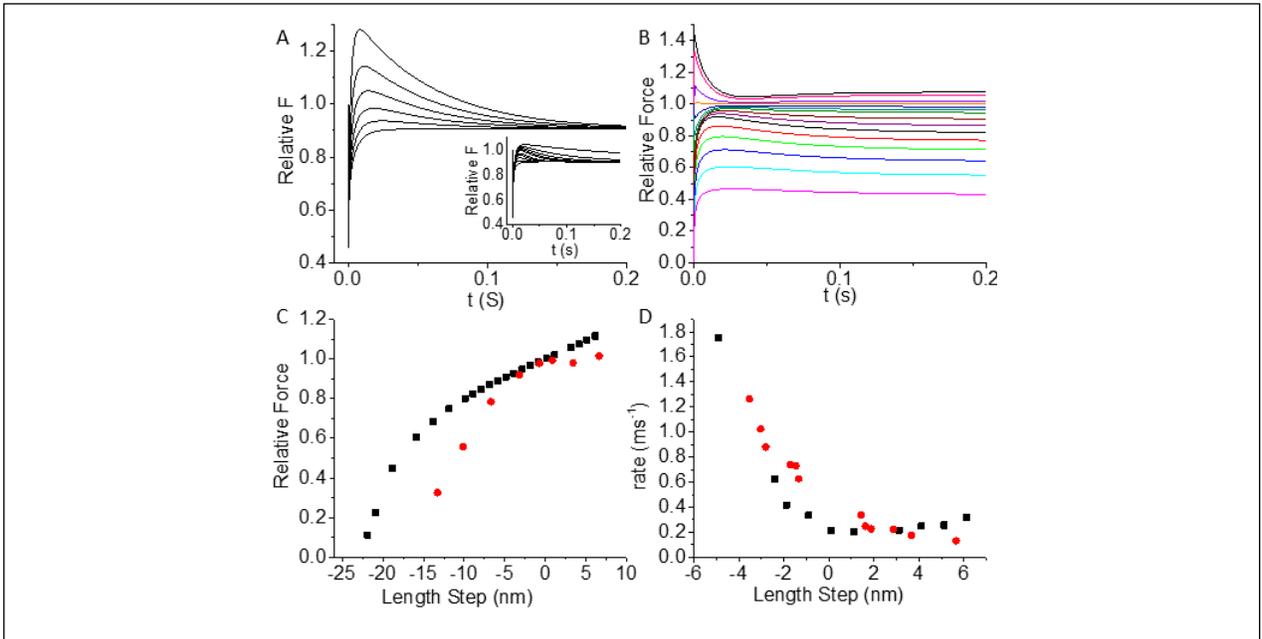

**Figure 4.** Simulated force transients following length steps, $\Delta L$, using three master equations (Eqs. 2, 3, and 5) and the rate constants in Eq. 4. Parameters were initialized to values in Table 1, and $N_{AMD} = N_{MDP} = 25$ and $F_i = 137.6$ pN (Relative $F_i = 1$) consistent with Eq. 1. (A) At t = 0, $F_i/\kappa_{sys}$ was decreased by $\Delta L = 4.89$ nm from a relative $F_i$ of 1 to 0.46. Force transients were simulated with an equilibration rate, $b$, of 20 s$^{-1}$ for different extents of adiabatic relaxation (0, 0.2, 0.4, 0.6, 0.8, and 1) where the bottom curve is non-adiabatic and the top curve is 100% adiabatic. (Inset) Force transients were simulated with 50% adiabatic chemical relaxation and $b$ values ranging from 0 to 240 s$^{-1}$. (B) Force transients were simulated as in panel A with 50% of $\Delta G$ damped and $b = 20$ s$^{-1}$ at different length steps, $\Delta L$. (C) Phase 2 amplitudes were determined from simulations in panel B, plotted (black squares) for different steps $\Delta L$ and overlayed (red circles) with experimental data (25) (D) Phase 2 rates were determined from simulated transients in panel B as $1/t_{½}$, plotted (black squares) for different $\Delta L$ and overlayed (red circles) with experimental data (25).

**[P$_i$] Step Simulations**. As shown by Dantzig et al. (27) a rapid increase in [P$_i$] upon photo-release of a caged-P$_i$ compound results in a multi-phasic force response. In these experiments, different final phosphate concentrations, [P$_i$]$_f$, were achieved through a combination of varying both initial concentrations, [P$_i$]$_i$, and the amount of caged-P$_i$ photo-released. The simulations below mimic these experiments.

The myosin working step is associated with the release of P$_i$ (13, 18, 28). Thus, a rapid increase in [P$_i$] from an initial concentration, [P$_i$]$_i$, to a final concentration, [P$_i$]$_f$, increases the system free energy by $\delta E = kT \cdot \ln([P_i]_f/[P_i]_i)$. Starting from the same initial equilibrium conditions (see Fig. 5 legend) used for the length-step experiments above (only here $f\_^\circ = 0.01$ s$^{-1}$) phase 1 is simulated simply by setting $\delta E$ to $kT \cdot \ln([P_i]_f/[P_i]_i)$ and $\Delta G(t)$ to $\Delta_r G$. Unlike in length step simulations (Fig. 5), force does not change with this transition, and so phase 1 is not mechanically observed.

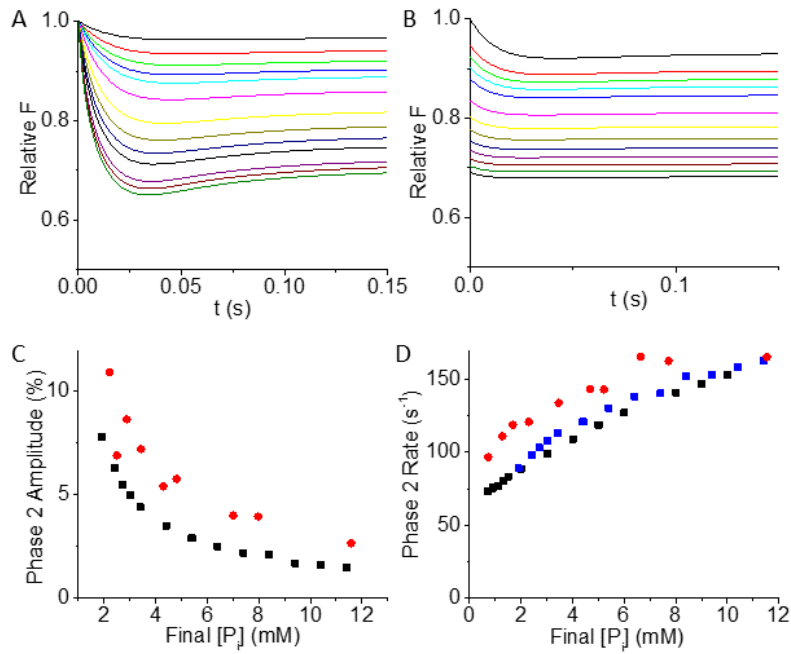

**Figure 5.** Simulated force transients following a rapid increase in [P$_i$] using three master equations (Eqs. 2, 3 and 5) and the rate constants in Eq. 4. Parameters were initialized to values in Table 1 (except $f\_^\circ = 0.01$ s$^{-1}$), $N_{AMD} = N_{MDP} = 25$, and $F_i = 137.6$ pN (Relative $F_i = 1$) consistent with Eq. 1. Additional assumptions are 50% adiabatic relaxation and that [P$_i$] affects only the reverse working step rate ($a_{dE} = 0$). (A) At t = 0, $\delta E = kT \cdot \ln($[P$_i$]$_f/$[P$_i$]$_i$ was added to Eq. 4. Force transients were simulated for an initial concentration, [P$_i$]$_i$, of 0.5 mM and final concentrations, [P$_i$]$_f$, ranging from 0.7 to 10 mM (curves top to bottom). (B) Force transients were simulated for initial concentrations, [P$_i$]$_i$, ranging from 0.5 to 10 mM (curves top to bottom) and final concentrations [P$_i$]$_f$ = [P$_i$]$_i$ + 1.4 mM P$_i$. (C) The percent decrease in force with phase 2 in panel B plotted at different [P$_i$]$_f$ (black squares) and overlayed with experimental data (red circles) from Dantzig et al. (27). (D) The phase 2 rate determined from single exponential fits of phase 2 in both panels A (blue squares) and B (black squares) is plotted and overlayed with experimental data (red circles) from Dantzig et al. (37). In neither panels C nor D are simulated transients best fits to experimental data.

Figures 5A and 5B show simulated chemical relaxations for different combinations of $[P_i]_f$ and $[P_i]_i$. Like in length step simulations, two phases (phases 2 and 3) emerge. Phase 2 occurs with changes in $N_{AMD}$, $N_{MDP}$, and $F$ in a direction that is consistent with Le Chatellier's principle and that minimizes the system free energy, approaching non-equilibrated values defined at $\Delta G(t)$ that overshoot values defined at $\Delta G = 0$. Phase 3 occurs upon equilibration when $N_{AMD}$, $N_{MDP}$, and $F$ change in a direction opposite phase 2 as $\Delta G(t)$ approaches 0 (Eq. 5). The new equilibrium values (relative to the equilibrium values prior to the length step) demonstrate that adding energy to the system, $\delta E$, through a rapid increase in $[P_i]$ creates a new isotherm. Figure 5C is a plot of phase 2 amplitudes obtained at different $[P_i]_f$ overlaid with experimental data (27). Figure 5D is a plot of phase 2 rates obtained at different $[P_i]_f$ overlaid with experimental data (27).

Considering that the above simulations are not best fits to experimental data, a binary mechanical model accounts reasonably well for the phases of a muscle tension response observed both following a rapid increase in $[P_i]$ (Fig. 5) and following rapid changes in muscle length (Fig. 4).

DISCUSSION

The force response to a rapid mechanical or chemical perturbation of muscle provides insights into basic relationships between actin-myosin kinetics, energetics, and force generation (25–27, 29, 30). The observed multi-phasic mechanical transients reveal fast kinetic and mechanical signatures of muscle that place significant constraints on models of muscle contraction. For example, according to a molecular mechanic model (5, 19), a discrete myosin working step is insufficient to account for muscle force transients, which, it has been argued, is evidence for additional force generating biochemical states and mechanisms (25, 27, 30). It could also be argued, however, that the inconsistency between this analysis and the direct observation of a single working step (Fig. 1A) implies that the molecular mechanic model does not accurately describe muscle force transients.

Here I have shown that a binary mechanical model system is sufficient to account for all four phases of a muscle force transient. While biochemical substates within the two mechanical states (not considered here) certainly influence the rates and amplitudes of the four phases, the four phases are inherently defined by and intrinsically emerge from a binary mechanical model system.

Molecular mechanic models provide the foundation for most models of muscle and molecular motor ensembles to date (5, 19). These models assume that molecular springs are defined mechanically not thermally, resulting in energy diagrams that describe the energy of individual motors rather than the free energy of a system of motors (Fig. 2). For example, a stress-strain curve of a molecular spring is the molecular mechanic equivalent of a reaction isotherm, where the elastic potential energy of a molecule is transferred to work through a mechanical relaxation rather than the system free energy being transferred to work through a chemical relaxation (Fig. 2).

A molecular mechanic model in effect formally projects system properties (Fig. 2) onto a molecule within which system properties do not physically exist. For example, the system force is projected onto a motor as an average force, and a molecular spring is then defined to contain that force and to use it circularly (top-down to bottom-up) to define the system force. This might be a useful exercise for imagining molecular mechanisms of system behaviors were it not for the fact that there are system properties other than force that cannot be projected on to a molecule. Unlike the thermally insulated molecular spring defined to formally contain muscle force within a molecule, there are no comparable molecular vessels for containing a reaction isotherm (phase 1), adiabatic chemical relaxation (phase 2), or equilibration (phase 3). Thus, a molecular mechanic analysis of force transients requires additional states and mechanisms to mimic these system properties, an exercise Gibbs described as seeking "mechanical definitions of temperature and entropy" (31).

While a binary mechanical model system may not provide easily accessible molecular images of the mechanisms underlying the system's mechanical performance, emergent thermodynamic mechanisms are physical, and molecular mechanic states and mechanisms required to mimic them are not. A binary mechanical model system is a novel and intriguing problem in statistical mechanics with potentially significant biomechanical applications.

ACKNOWLEDGEMENTS

I thank Willard, AV, Julie and my students, colleagues, and mentors who have over many years inspired and guided this work. This work was funded by a grant from the National Institutes of Health 1R01HL090938-01.

**References**

1. Kittel, C., and H. Kroemer. 1980. Thermal Physics. .

2. Baker, J.E., and D.D. Thomas. 2000. A thermodynamic muscle model and a chemical basis for A . V . Hill ' s muscle equation. *J Muscle Res Cell Motil.* 21:335–344.

3. Baker, J.E. 2022. A Chemical Thermodynamic Model of Motor Enzymes Reconciles Differences Between Chemical-Fx and Powerstroke Models,. *Biophys. J.*

4. Hill, A.V. 1938. The heat of shortening and the dynamic constants of muscle. *Proc. R. Soc. London. Ser. B*. 126:136–195.

5. Huxley, A.F. 1957. Muscle structure and theories of contraction. *Prog. Biophys. Biophys. Chem.* 7:255–318.

6. Baker, J.E., I. Brust-Mascher, S. Ramachandran, L.E. LaConte, and D.D. Thomas. 1998. A large and distinct rotation of the myosin light chain domain occurs upon muscle contraction. *Proc. Natl. Acad. Sci. U. S. A.* 95:2944–9.

7. Sweeney, H.L., A. Houdusse, and J. Robert-Paganin. 2020. Myosin Structures. In: Caluccio LM, editor. Myosins, Advances in Experimental Medicine and Biology. Springer Nature Switzerland AG. pp. 7–19.

8. Rayment, I., W.R. Rypniewski, K. Schmidt-Bäse, R. Smith, D.R. Tomchick, M.M. Benning, D.A. Winkelmann, G. Wesenberg, and H.M. Holden. 1993. Three-dimensional structure of myosin subfragment-1: A molecular motor. *Science (80-. ).* 261:50–8.

9. Huxley, H.E. 1969. The mechanism of muscular contraction. *Science (80-. ).* 164:1356–65.

10. Sabido-David, C., S.C. Hopkins, L.D. Saraswat, S. Lowey, Y.E. Goldman, and M. Irving. 1998. Orientation changes of fluorescent probes at five sites on the myosin regulatory light chain during contraction of single skeletal muscle fibres. *J. Mol. Biol.* 279:387–402.

11. Huxley, A.F. 1974. Muscular contraction. *J. Physiol.* 243:1–43.

12. Uyeda, T.Q., P.D. Abramson, and J.A. Spudich. 1996. The neck region of the myosin motor domain acts as a lever arm to generate movement. *Proc. Natl. Acad. Sci. U. S. A.* 93:4459–64.

13. Baker, J.E., C. Brosseau, P.B. Joel, and D.M. Warshaw. 2002. The biochemical kinetics underlying actin movement generated by one and many skeletal muscle myosin molecules. *Biophys. J.* 82:2134–47.

14. Hooft, A.M., E.J. Maki, K.K. Cox, and J.E. Baker. 2007. An accelerated state of myosin-based actin motility. *Biochemistry*. 46:3513–20.

15. Stewart, T.T.J., D.R. Jackson, R.D.R. Smith, S.F.S. Shannon, C.R. Cremo, and J.E. Baker. 2013. Actin Sliding Velocities are Influenced by the Driving Forces of Actin-Myosin Binding. *Cell. Mol. Bioeng.* 6:26–37.

16. Brizendine, R.K., G.G. Sheehy, D.B. Alcala, S.I. Novenschi, J.E. Baker, and C.R. Cremo. 2017. A mixed-kinetic model describes unloaded velocities of smooth, skeletal, and cardiac muscle myosin filaments in vitro. *Sci. Adv.* 3.

17. Stewart, T.J., V. Murthy, S.P. Dugan, and J.E. Baker. 2021. Velocity of myosin-based actin sliding depends on attachment and detachment kinetics and reaches a maximum when myosin binding sites on actin saturate. *J. Biol. Chem.* 101178.

18. Baker, J.E., L.E.W. LaConte, I. Brust-Mascher, and D.D. Thomas. 1999. Mechanochemical coupling in spin-labeled, active, isometric muscle. *Biophys. J.* 77:2657–64.

19. Hill, T.L. 1974. Theoretical formalism for the sliding filament model of contraction of

striated muscle. Part I. *Prog. Biophys. Mol. Biol.* 28:267–340.

20. Lymn, R.W., and E.W. Taylor. 1971. Mechanism of adenosine triphosphate hydrolysis by actomyosin. *Biochemistry*. 10:4617–4624.

21. Goldman, Y.E. 1987. Kinetics of the actomyosin ATPase in muscle fibers. *Annu. Rev. Physiol.* 49:637–54.

22. Cooke, R. 1997. Actomyosin interaction in striated muscle. *Physiol. Rev.* 77:671–97.

23. Gasser, H.S., and A.V. Hill. 1924. The dynamics of muscular contraction. *Proc R Soc L. B Biol Sci*. 398–437.

24. Jewell, B.R., and D.R. Wilkie. 1958. An analysis of the mechanical components in frog's striated muscle. *J. Physiol.* 143:515–540.

25. Huxley, A.F., and R.M. Simmons. 1971. Proposed mechanism of force generation in striated muscle. *Nature*. 233:533–538.

26. Ford, L.E., A.F. Huxley, and R.M. Simmons. 1977. Tension responses to sudden length change in stimulated frog muscle fibres near slack length. 269:441–515.

27. Dantzig, J., Y. Goldman, N. Millar, J. Lacktis, and E. Homsher. 1992. Reversal of the cross-bridge force-generating transition by photogeneration of phosphate in rabbit psoas muscle fibres. *J. Physiol.* 451:247.

28. Cooke, R., and E. Pate. 1985. The effects of ADP and phosphate on the contraction of muscle fibers. *Biophys. J.* 48:789–98.

29. Civan, M.M., and R.J. Podolsky. 1966. Contraction kinetics of striated muscle fibres following quick changes in load. *J. Physiol*. 511–534.

30. Kawai, M., and H.R. Halvorson. 1991. Two step mechanism of phosphate release and the

mechanism of force generation in chemically skinned fibers of rabbit psoas muscle. *Biophys. J.*

31. Gibbs, J. 1902. Elementary Principles in Statistical Mechanics Developed with Especial Reference to the Rational Foundation of Thermodynamics. .

32. Guilford, W.H.H., D.E.E. Dupuis, G. Kennedy, J. Wu, J.B.B. Patlak, and D.M.M. Warshaw. 1997. Smooth muscle and skeletal muscle myosins produce similar unitary forces and displacements in the laser trap. *Biophys. J.* 72:1006–21.

33. Ford, L.E., A.F. Huxley, and R.M. Simmons. 1981. The relation between stiffness and filament overlap in stimulated frog muscle fibres. *J. Physiol.* 311:219–249.

34. Lewalle, A., W. Steffen, O. Stevenson, Z. Ouyang, and J. Sleep. 2008. Single-molecule measurement of the stiffness of the rigor myosin head. *Biophys. J.* 94:2160–9.

35. Pertici, I., G. Bianchi, L. Bongini, D. Cojoc, M.H. Taft, D.J. Manstein, V. Lombardi, and P. Bianco. 2021. Muscle myosin performance measured with a synthetic nanomachine reveals a class-specific Ca2+-sensitivity of the frog myosin II isoform. *J. Physiol.* 599:1815–1831.

36. Veigel, C., J.E. Molloy, S. Schmitz, and J. Kendrick-jones. 2003. Load-dependent kinetics of force production by smooth muscle myosin measured with optical tweezers. *Nat. Cell Biol.* 5:980–986.

37. Dantzig, J., M. Hibberd, D. Trentham, and Y. Goldman. 1991. Cross-bridge kinetics in the presence of MgADP investigated by photolysis of caged ATP in rabbit psoas muscle fibres. *J. Physiol.* 432:639.

**Table 1.** Model Parameters. All values are consistent with experimental measurements in muscle and muscle myosin where rate constant references are from R temporaria and mechanical parameters references are from skeletal muscle. The parameter $a_{Go}$ was determined from $f_+^o = \exp[a_{Go} \cdot \Delta G^o/kT]$ assuming $f_+^o = 40$ s$^{-1}$ and gives a value for $f_-^o = \exp[-(1 - a_{Go}) \cdot \Delta G^o/kT]$ of 0.1 s$^{-1}$. All other model parameters are determined using first principles (equations derived herein).

| Parameter | Description | Value |
|---|---|---|
| N | Number of myosin heads | 50 (17) |
| $\Delta G^o$ | Standard reaction free energy for working step | 6 kT (2) |
| d | Step size | 8.7 nm (32) |
| $\kappa_{sys}$ | Effective stiffness of system spring | 15.5 pN/nm (33, 34) |
| y | Fraction of force generating myosin heads | 0.3 |
| b | Rate of equilibration | 20 s$^{-1}$ |
| $a_{Go}$ | partition of standard reaction free energy | 0.615 (35) |
| $a_{Fd}$ | partition of work | 0.3 |